\documentstyle{aipproc}

\begin{document}

\title{HETE-II and the Interplanetary Network}
\author{K. Hurley$^1$, J.-L. Atteia$^2$,G. Crew, G. Ricker, J. Doty, G. Monnelly, R. Vanderspek, J. Villasenor$^3$, T. Cline$^4$}
\address{$^1$UC Berkeley Space Sciences Laboratory, Berkeley, CA 94720-7450\\
$^2$CESR, BP 4346, 31028 Toulouse Cedex 4, France\\
$^3$M.I.T. Center for Space Research, 70 Vassar St., Cambridge, MA 02139\\
$^4$NASA GSFC, Code 661, Greenbelt, MD 20771}

\maketitle

\begin{abstract}

The FREGATE experiment aboard HETE-II has been successfully integrated into the
Third Interplanetary Network (IPN) of gamma-ray burst detectors.  We show how HETE's
timing has been verified in flight, and discuss what HETE can do for the IPN and vice-versa.

\end{abstract}

\section{Introduction}

The FREGATE experiment aboard HETE-II is an excellent complement to the detectors in
the IPN.  It has good sensitivity to bursts, thanks to its large surface area and its
steady background in equatorial orbit.  Its time resolution in both triggered and untriggered modes
is sufficiently high for cross-correlating precisely with other spacecraft.  And finally, its energy range is well
matched to those of other instruments in the network.  However, before any experiment can be
added to the network, its timing must be verified in flight.  In this paper, we will first explain the
procedure we have used to demonstrate
that the HETE timing is accurate.  Then we will show how the IPN can be used to improve the
location accuracy of HETE bursts, and how the FREGATE data are useful to the IPN.  

\section{Timing}

In principle, we expect the timing of the HETE spacecraft to be good, both because it is
in low Earth orbit, and because it is derived from an onboard GPS receiver.  While there
are several ways to confirm the timing of any spacecraft in flight, the best is to triangulate
burst sources with known positions, either soft gamma repeaters, or gamma-ray bursts
for which optical or radio counterparts are found.  In its first year of operation, FREGATE
detected numerous bursts from SGR1806-20 and 1900+14; seven of them were observed
in conjunction with other spacecraft in the IPN.  The positions of both SGRs are known to
arcsecond or sub-arcsecond accuracy, and therefore serve as timing calibration sources.

In figures 1 and 2 we show HETE/Ulysses annuli for bursts from each SGR.  From these
triangulations we conclude that HETE timing errors are indeed negligible, and moreover,
that HETE is a sensitive SGR detector when it is pointed towards the Galactic plane.  Since this
occurs for several months around December and June, it is clear that HETE will serve
as a monitor of the known SGRs, and may also detect new ones when they become
active.  

\section{What HETE brings to the IPN}

At present, the IPN consists of Ulysses, in heliocentric orbit, Mars Odyssey, now in orbit
around Mars, and several missions close to Earth (Cline et al. 2002).  Prior to the launch of HETE, the two principal GRB detectors 
close to Earth were the BeppoSAX Gamma-
Ray Burst Monitor (GRBM) and the Konus experiment aboard the Wind spacecraft.  BeppoSAX has a high
duty cycle, but as it is in low-Earth orbit, it misses roughly half of all GRBs due to Earth occultation.  The GRBM
has a coarse location capability (10-20 $^o$).  The mission will end some time in 2002.
Wind is nominally at the L$_1$ Lagrange point, so
no bursts are missed due to occultation, but the duty cycle is not as high as that of BeppoSAX, and this causes
some events to be missed.  Because of the placement of the two Konus detectors aboard the spacecraft, the
experiment can be used to determine the ecliptic latitude of a burst to an accuracy of about 10$^o$ in many,
but not all, cases.  The Wind mission is now expected to be supported at least through 2002, at which time
the launch of INTEGRAL should provide a replacement or near-Earth vertex for the IPN.

HETE complements these missions in three ways, as far as the IPN is concerned.
First, it fill in the gaps by providing data on bursts missed by the
other spacecraft.  To date, about one burst per month has been detected by FREGATE and verified
by Ulysses or Mars Odyssey, but has
not been detected by Konus or by the GRBM.  Second, it provides burst data in near-real time.  The GRBM data
are delayed by up to several hours, and the Konus data by half a day or more.  Thus even when
these experiments detect a burst, valuable time can be gained by receiving the HETE data and
alerting the Ulysses and Mars Odyssey teams at JPL to expedite processing of the data in the
small time window where the burst should occur.  Finally, FREGATE data can be used to resolve
the ambiguity in a 3 or 4 spacecraft localization.  In the case where Ulysses, Mars Odyssey, and
FREGATE observe a burst, FREGATE's coarse localization capability can serve to indicate
which of two alternate triangulation positions is the correct one.  In the case where Ulysses, Mars Odyssey,
FREGATE, and Konus observe a burst, the separation between FREGATE and Konus will often
be just enough to generate a third, independent annulus of location which again resolves the
ambiguity. It should also be noted that the time resolution of FREGATE in the untriggered
mode is 0.164 s, which is a factor of about 5 to 20 better than the untriggered modes of
the GRBM and Konus, which results in more accurate cross-correlations.

\section{What the IPN does for HETE}

The IPN can assist the HETE mission in four ways.  First, it can confirm candidate
GRBs and SGR bursts.  Typically, these are events which are observed only by
FREGATE.  24 such bursts have been confirmed to date, or a rate of about two
per month.

Second, it can refine WXM error circles.  Figure 3 shows one example, GRB010613.
This is an extreme case, in that both the WXM error circle and the IPN annulus
are rather large.  However, there are numerous examples of smaller BeppoSAX WFC
error circles which were refined by the IPN (Hurley et al. 2001), and we expect most
HETE-IPN bursts to be similar to them.

Third, it can transform "one dimensional" WXM error boxes which are too
large to search for counterparts into small error boxes
which are suitable for optical and radio follow-up observations.  A "one-dimensional"
WXM error box is one which results when a burst is detected and localized in only one of the two
crossed WXM detectors, leading to a long, narrow error box.  Figure 4 shows an example,
GRB010921.  Figure 5 shows an enlargement of the region where the IPN annulus intersects
the WXM error box.  This burst and its counterpart are described in detail elsewhere
(Ricker et al. 2002, Park et al. 2002, Kulkarni et al. 2002).

Finally, the IPN can, in principle, transform one dimensional SXC error boxes
into small error boxes in the same way that it transforms WXM error boxes.
However, this has not yet been done for any burst, since none has been localized by
the SXC in this way.
 
\section{Conclusions}

We have successfully integrated the HETE spacecraft into the third interplanetary network of gamma-ray
burst detectors, and have demonstrated the value of HETE to the IPN and vice-versa.  We are now looking
forward to several more years of operations.

KH is grateful for IPN support under JPL Contract 958056, and for HETE-FREGATE
support under MIT Contract SC-R-293291.

\pagebreak

\begin{figure}
\special{gif:figure1.gif}
\caption{The FREGATE/Ulysses location annulus for a burst from SGR1900+14 on June 29 2001.
The annulus width is 11'.  The displacement between the known position of the SGR and the
center line of the annulus is 31", which corresponds to a timing error of 93 ms.  As this is much
less than the 1 sigma statistical uncertainty in the cross-correlation, it indicates that the HETE timing
errors are negligible.}
\label{fig1}
\end{figure}

\begin{figure}
\special{gif:figure2.gif}
\caption{The FREGATE/Ulysses location annulus for a burst from SGR1806-20 on June 23 2001.
The annulus width is 12'.  The displacement between the known position of the SGR and the
center line of the annulus is 1", which corresponds to a negligible timing error of 3 ms.}
\label{fig2}
\end{figure}

\begin{figure}
\special{gif:figure3.gif}
\caption{The 72' diameter WXM error circle for GRB010613, and the
15' wide FREGATE/Ulysses location annulus for it.  The intersection of the
annulus and the error circle forms an error box whose area is approximately 1000 square
arcminutes, or about a factor of four smaller than the error circle alone.  A similar
annulus could also have been generated using the Konus-Wind data, had FREGATE
not detected this burst for any reason.}
\label{fig3}
\end{figure}

\begin{figure}
\special{gif:figure4.gif}
\caption{The 5.2$^o$ by 17' one-dimensional WXM error box for GRB010921, and the
14' wide FREGATE/Ulysses location annulus for it.  The position of the optical transient
is marked with an asterisk. A similar
annulus could also have been generated using the Konus-Wind data, had FREGATE
not detected this burst for any reason.}
\label{fig4}
\end{figure}

\begin{figure}
\special{gif:figure5.gif}
\caption{Expanded view of the 310 square arcminute IPN/WXM error box for GRB010921, and the
position of the optical transient.}
\label{fig5}
\end{figure}

\end{document}